\documentclass[prd,aps,twocolumn,floatfix]{revtex4}
\usepackage{amssymb,graphicx,epsfig}

\begin{document}

\title{Dynamical control of the constraints growth in free
  evolutions of Einstein's equations.}

\author{Manuel Tiglio}

\affiliation{Department of Physics and Astronomy, Louisiana State
University, 202 Nicholson Hall, Baton Rouge, LA 70803-4001}

\begin{abstract}
I present a new, simple method to dynamically control the growth of
the discretized
constraints during a free evolution of Einstein's
equations. During an 
evolution, any given family of formulations is adjusted off the constraints surface in a
way such that, for any chosen numerical method and arbitrary but fixed
resolution, 
the constraints growth
can be minimized with respect to the freedom allowed by the 
formulation. In particular, provided there is enough freedom, the
discretized constraints can be maintained close to its
initial truncation
value for all times, or decay from it. 

No a priori knowledge of the solution
is needed, and the method can be applied to any formulation of Einstein's
equations without affecting hyperbolicity. This method is independent
of the numerical algorithm   
and accounts for constraint violating modes
introduced both by continuum instabilities of the formulation and by the
numerical method. 

\end{abstract}
\maketitle

\section{Motivation and overview}

The Einstein equations are typically solved in 
what is called free or unconstrained evolutions. The equations are
split into a set of evolution equations and another one of
constraints. The standard procedure is to use initial data that
satisfy the constraints and  later solve only for the set of evolution
equations. 
The logic is that by virtue of Bianchi identities the
constraints should be satisfied at all times if they do so initially
and the evolution equations are solved. This is actually true only in
the absence of boundaries, or in the future domain of dependence of the
initial data; in the presence of boundaries appropriate boundary
conditions have to given in such a way that the associated
initial-boundary value problem is well posed, a highly non-trivial
issue on its own (see \cite{boundaries} for
related work). 

However, problems arise in free evolutions even neglecting the
presence of boundaries. Although the constraints should be
exactly satisfied at the continuum level, in typical, fixed
resolution, 
numerical simulations of strong fields the constraints quickly grow 
and eventually the code crashes. For a numerically 
stable scheme (throughout this paper,
the term {\it numerical stability} is used as equivalent to
convergence, in the sense of Lax's theorem \footnote{That is, the
  existence of a precise bound for the numerical 
solution in terms of the initial data, with the bound
being independent of resolution.}) this growth should
go away with resolution, but very high resolutions are usually 
needed if one wants to keep these errors under control by just adding more points to the simulation. Therefore this procedure is
not practical and it might not be even feasible.

Sometimes it happens there are no
growing modes at the continuum, but the
discretization one is using, even if numerically stable, can introduce
errors that grow quickly in time. Examples of this potential source of
 instabilities are shown in
Ref. \cite{exc}; it is there also explained how to
rearrange the semidiscrete equations in order to prevent this when there
is a ``conserved'' quantity at the continuum (for example, in the case
of wave propagation on a stationary spacetime the physical energy would play this
role). 

The picture that has emerged in other cases is that rapidly growing
discrete constraint violating modes are numerical excitations of 
unstable continuum modes. In other words, solutions of 
Einstein's evolution equations that are initially slightly off
the constraint surface, but then deviate from it very quickly. The
standard approach is then to seek formulations that are as stable as
possible under continuum constraint violations. A trivial
observation here is that when rapidly growing discrete constraint
violations appear 
it is because the given numerical method allows for them.  

Therefore, I
argue that 
it is desirable to control the {\em discretized}
constraints, taking into account possible growth introduced both by the
formulation of the equations, {\em and} by the numerical method.  The
purpose of this paper is to present a method for doing so. 

As one is interested in solutions that satisfy the constraints,
the Einstein evolution equations can be written in infinitely
different ways by changing the system's behavior off of the constraint surface. For
definiteness I will now concentrate on first order (both in space and
time) formulations, though the ideas of this paper clearly do not
depend on this and can be applied to second order formulations as
well. Beginning with any
formulation, e.g. a  symmetric hyperbolic one, 
one can add to the right hand
side (RHS) terms proportional to the constraints. That is
\begin{equation}
\dot{u} = \sum_jA^j(u,t,\vec{x})\partial_ju + B(u,t,\vec{x}) + \mu C \label{linearc}
\end{equation}
where $B$ and $u$ are a vector valued functions (containing the metric and
related variables), 
\begin{equation}
C=\left(C_1 \ldots C_n \right)\;,  \label{constraints}
\end{equation}
with $ C_i = C_i(u,\partial_ju) $, is a vector valued constraint (containing the
Hamiltonian, momentum, and perhaps other non-physical constraints that appear
as integrability conditions when the system is written in first order form),
and $A,\mu$ are matrix-valued functions. The constraints
(\ref{constraints}) in Einstein's equations are quasilinear. That is, nonlinear in the field
variables $u$, but linear in the spatial derivatives, $\partial_i u$. 
 Adding the constraints in a linear
way, as in equation (\ref{linearc}), yields a quasilinear 
system of hyperbolic partial differential equations, for which many
results from the mathematical literature apply \footnote{During preparation of this
  manuscript, related by work by David Fiske appeared \cite{fiske}, in
  which the 
  motivation is to make the constraints decay exponentially to zero for any
spacetime (i.e. without any a priori knowledge). Presumably the
resulting equations are not quasilinear (or linearly degenerate in
the vacuum case).}.

The freedom in choosing $\mu$
was originally used to cast Einstein's equations in symmetric hyperbolic
form, later to obtain ``physical'' characteristic speeds (see
\cite{hyp} for reviews), and more recently to improve the stability properties of formulations
around fixed backgrounds (the number of papers in the area is too large to be
reviewed here; see \cite{stability} and references therein). Typically $\mu$ is chosen as a
constant matrix; however, it does not appear to be the most effective way of controlling
deviations off the constraint surface in a generic case for two reasons: i) some
information about the solution (typically, the assumption that it will be
close -or actually identical - to  some known
background) is usually needed in order to choose an optimal
$\mu$. However, this kind of information will generically not be
available. ii) In principle, constants 
do not seem the best way to control varying fields. 

Choosing constant values for $\mu$ seems to be favored for historical
reasons, and it is actually not required. Symmetric hyperbolicity and 
physical characteristic speeds, for example, can be achieved even if
 using functions of spacetime, provided they are a
priori given. See \cite{st} for one example. 

\section{Dynamical minimization of the constraints growth: the main idea}

Consider the constraints $C=C(u,D_iu)$, where the operator $D$ may be 
either a continuum or discrete derivative. Later $D$ will represent
discrete derivatives, since the motivation is to control not only
continuum constraint violations, but also violations that arise in the
discrete equations. 
For any slicing of spacetime $S _t \times {\cal R}$, define a 
norm $N_c=N_c(t)$ for the
constraints. Note I will always refer to this norm, not the norm
 associated with the main field variables (will comment on this
in the last section). For simplicity I choose the $L_2$ norm,
$$
N_c(t) = \frac{1}{2}\int_{S_t} \sum_iC_i^2
$$
where integration is on the spatial hypersurface $S_t$ \footnote{If the
derivative is discrete this is to be replaced by a discrete sum, of course.}. 

One wants this norm either not to grow during an evolution, or least 
to grow as slowly as possible. As $C$ is a function of the field variables, 
any solution to the evolution equations automatically determines the
rate growth for $N_c$, independent of whether one is considering
the fully discrete, semidiscrete or continuum systems. I will concentrate
on the last two cases (i.e., with continuous time). In principle the dependence of $N_c$ on the
evolution equations is complicated. However, given that $\mu $ is added 
linearly to the right hand side of the evolution equations, as in 
Eq.\ref{linearc}, the time derivative of $N_c$ can be written as
\begin{eqnarray*}
\dot{N}_c &=& \int_{S_t} \sum_iC_i\dot{C_i} \\
& =& \int_{S_t} \sum_{i,j} C_i\left[ \frac{\partial C_i}{\partial u_j}\dot{u_j}
 + \sum_k\frac{\partial C_i}{\partial (D_k u_j)} D_k\dot{u_j}
\right]
\end{eqnarray*}
Using the evolution equations Eq.(\ref{linearc}) and allowing $\mu $ to depend
only on time (not on space), the time derivative of the norm can then be written as
\begin{equation}
\dot{N}_c = I^{(1)} + \mbox{trace} (\mu \times  I^{(2)}) \label{split}
\end{equation}
where
\begin{eqnarray}
 I^{(1)} &=&
 \int\sum_{i,j} C_i\left[\frac{\partial C_i}{\partial u_j}+
\sum_k\frac{\partial C_i}{\partial D_k u_j}D_k \right] \times
 \nonumber \\
&&  \left[\sum_l(A^lD_lu_j) +B_j\right]  \label{split1}  \\
I^{(2)}_{jl} &=& \int \sum_{i} C_i\left[\frac{\partial C_i}{\partial u_j}+
\sum_k\frac{\partial C_i}{\partial D_k u_j}D_k \right]C_l \label{split2}
\end{eqnarray}
The quantities $\{ N_c,I^{(1)},I^{(2)},\mu \}$ depend only on time, in the case of
$\mu$ by assumption and in the others because spatial integration
is carried out. The splitting
 (\ref{split}) is a crucial step in the method presented here,
since one gains analytical control on the norm growth as a function of
$\mu$. More explicitly: $\dot{N_c}$ is a linear function of $\mu$ (because the
 constraints are added linearly to the evolution equations).

The idea now is to choose $\mu$ such that $\dot{N_c}$ has some desired
behavior within
the freedom allowed by a given formulation of the evolution equations. 
As a simple example, consider a single scalar equation. One could
choose $\dot{N}_c=0$ by defining $\mu = -I^{(1)}/I^{(2)}$. Or one could
choose the norm to decay at a constant or exponential rate
 by defining $\mu = -(I^{(1)}+d^2)/I^{(2)}$ or 
$\mu =  -(I^{(1)}+d^2N_c)/I^{(2)}$, respectively, with $d$ some constant. 
On the other hand, if the different components of the $\mu$
matrix are restricted to some sets (this is a
restriction that appears quite often when requiring symmetric hyperbolicity, 
physical characteristic speeds, or both) one could define $\mu(t)$ 
as the one that gives the ``smallest'' growth of $\dot{N}_c(t)$ in those
sets, etc.  

This could be done at the continuum. However, the problem is that $\mu$ would
 then depend on the constraints, which in turn depend on the derivatives of
the field variables. As a consequence, the resulting evolution
equations would not be quasilinear, and it is far from clear that they would define
a well posed initial, or initial-boundary, value problem. Indeed,
in Section IV I present numerical evidence that strongly
suggests that the resulting system would be ill posed. 

Therefore, I propose to define $\mu $ through a single
resolution run and interpolate the obtained discrete function in order to have it 
defined at all times. In an actual computation one would choose 
a fixed resolution and dynamically compute $I^{(1)},I^{(2)}$ and thus a
 $\mu$ that gives the desired norm behavior, making sure
that the resulting $\mu(t)$ is within the range allowed by symmetric
hyperbolicity (symmetric hyperbolicity is not a requirement of the
method, but ensures well posedness of the initial-boundary value
problem, provided appropriate boundary conditions are given). 
In order to ensure numerical stability this $\mu $ matrix valued 
function has to be kept fixed at
other resolutions. $\mu$ will depend on the original resolution 
used to define it, but this is not a problem. The important thing is
to keep the constraints under control for a given resolution. They
will also remain under control for better resolutions if one
is in the convergence regime. 

$\mu $ will also depend on the given problem of interest, and on the 
numerical method used to solve the equations. That is, for any chosen 
desired behaviour for $\dot{N}_c$, there will be one $\mu$, and one well
posed formulation of the problem for each physical situation one wants
to solve for with a given numerical method. This is not a practical
problem and, in fact, it seems difficult to
find a formulation that has optimum stability properties for
all possible solutions and all possible numerical methods.

\section{Controlling the constraints norm growth: more details}

If possible, one might want the constraints norm growth to be identically zero at all times 
 $\dot{N}_c=0$. A potential problem with this approach is that the resulting
 $\mu$ might have large values, rapid variations in
 time, or both. The resulting set of PDE's would then be stiff, having
  equivalent short timescales, in which case a very small Courant
 factor would be required. In the next section I show an example
 where this happens and argue that better results may be obtained by
 not enforcing $\dot{N}_c=0$. Even if one has enough freedom to
 enforce $\dot{N}_c=0$,
it appears that allowing $N_c$ for some fluctuations around the initial
value gives evolutions with better stability. I now elaborate on one
 way of doing so, 
assuming one has enough freedom in $\mu$. This approach is 
highly non unique, and many variations of the fundamental idea presented
 in this paper seem possible, and worth further exploration. If one
chooses
\begin{equation} 
\dot{N_c} = -a N_c   \label{edot}
\end{equation}
with $a>0$, any violation in the constraints will decay exponentially 
\begin{equation}
N_c(t+\Delta t) = N_c(t)e^{-a\Delta t} \label{decay}
\end{equation}
Choosing a constant $a$ makes the constraints decay for all
times, I will discuss this in the last section. Here I take a
different approach. I choose a a tolerance value for the
 norm, $N_c=T$, and solve for $\mu$ such that the constraints decay to this 
tolerance value after a given relaxation time. 
More precisely, I choose $a$ such that after time $n_a \Delta t$  
the constraints will have the
value $T$. Replacing $N_c(t+\Delta t)$ by $T$ in equation (\ref{decay})
and solving for $a$ gives
\begin{equation}
a(t) = -\frac{1}{n_a\Delta t}\ln{\left(\frac{T}{N_c(t)}\right)} \; . \label{a}
\end{equation}
I now solve $\dot{N_c} = -a N_c = I^{(1)} + \mbox{trace}(\mu I^{(2)})$
for $\mu$. One solution (non unique, since the equation is scalar and
$\mu$ is a matrix) is, assuming $I^{(2)}$ is
invertible, and ommiting all the indexes, 
\begin{equation}
\mu = -\frac{\left[I^{(2)}\right]^{-1}\left(aN_c
+I^{(1)}\right)}{\dim{\left[I^{(2)}\right]}} \; , \label{mu}
\end{equation}
with $a(t)$ given by eq.(\ref{a}). Thus, for any value of
$N_c$ at time $t$, at time
$t+n_a\Delta t$ the value will be $T$. Later in this paper $\Delta t $ will be the
discrete time step and, in particular, if $n_a$=1,
then $N_c=T$ at each time step. 

It must be emphasized that these results hold in the semidiscrete
case. I.e., for {\em any} - not necessarily high - spatial
resolution. But since time has been assumed continuous, 
the Courant factor must be such that the fully
discrete simulation  faithfully represents the above semidiscrete
calculations (another option would be to perform a fully
discrete analysis). In the next section I present some numerical
experiments to 
study, among other issues, the extent to which the fully discrete system represents 
the  semidiscrete analysis. As we will see, at least in the
examples here considered, standard values of the
Courant factor already yield good results. 

\section{Some numerical examples}

In these numerical experiments I use the method of lines
and finite differencing. The difference operator chosen is the
simplest one that satisfies summation by parts (second order in the
interior, and first order in the boundaries), numerical dissipation is
introduced taken into account modifications at boundaries, 
and third order Runge Kutta is used as time integrator (see
\cite{exc}). 

\subsection{The equations}
I use the standard ADM equations in spherical symmetry 
with the exact lapse-area
locking choices for the lapse and shift \cite{area} to illustrate and
study the method  
presented here. This choice of lapse and shift amounts to 
specifying in an arbitrary, but
a priori, way the lapse as a function
of spacetime, and
the shift as $ \beta = \alpha r K_b$. This choice of shift  
 implies that the radial area does not change in time, 
$\dot{g}_{\theta \theta }=0$, and one can choose $r=g_{\theta
  \theta }^{1/2} $ as a coordinate. The 3-metric and extrinsic curvature in
coordinates $r,\theta, \phi$ then take the form 
$g_{ij}=diag(a^2,r^2,r^2)$, $K^i_{\;\;j} = diag(K_a,K_b,K_b)$. 

The evolution equations, with the product of the Hamiltonian
constraint and $\mu = \mu(t)$ added to the time derivative of the $K_a$ equation, are
\begin{eqnarray}
\dot{a} &=& \alpha r(aK_b)' + a\left[\alpha (K_b-K_a) + rK_b \alpha
'\right] \label{adot}\\
\dot{K_a} &=& (\alpha 'r + 2\alpha)a^{-3}r^{-1}a'  - a^{-2}\alpha^{''}
+ \nonumber \\ 
& & \alpha\left[rK_bK_a' + K_a(K_a+2K_b) \right] + \mu
H \label{kadot} \\
\dot{K_b} &=& \alpha \left(a^{-3}ra' + rK_bK_b'  \right)
-a^{-2}r^{-1}\alpha ' + \nonumber \\
& & \alpha \left[K_b(K_a+2K_b)+r^{-2}(1-a^{-2})\right] \label{kbdot}
\end{eqnarray}
These equations constitute a strongly hyperbolic system, and the characteristic
speeds are ``physical''(along the light cone or normal to the spatial
hypersurfaces): $\beta , \beta \pm \alpha/a$, for any function $\mu
(t)$. That is, the equations are strongly hyperbolic even for $\mu =0$, i.e.,
the standard ADM equations in spherical symmetry with this choice
of gauge are strongly hyperbolic. 

The Hamiltonian and momentum constraints are, respectively,
\begin{eqnarray}
H&=& \frac{4}{ra^3}a' + \frac{2}{r^2}(1-a^{-2}) +  
 2K_b(K_b + 2K_a)  \label{ham}\\
M&=& K_b' + \frac{K_b-K_a}{r} \label{mom}
\end{eqnarray}
I perform these experiments with the  Painlev\'e-Gullstrand (PG) slicing of
the Schwarzschild black hole spacetime,
\begin{equation}
a=1 \;\;\; , \;\;\; 
K_a = -\frac{\beta }{2r} \;\;\; , \;\;\;
K_b = \frac{\beta}{r}
\end{equation}
\begin{equation}
\alpha = 1 \;\;\; , \;\;\;  \beta = \alpha r K_b \label{gauge}
\end{equation}
This is a stationary, exact solution, of the evolution and constraint
equations, 
(\ref{adot},\ref{kadot},\ref{kbdot}), and (\ref{ham},\ref{mom}),
respectively. 

As boundary conditions I set the time derivative of the incoming
characteristic modes to
zero. This will define a well posed initial-boundary value problem
but in general will violate the constraints. One possibility would
be to derive constraint-preserving boundary conditions for this
problem. However, this is not the purpose of this paper but, rather, to 
devise a mechanism to control the
constraint growth having fixed, among other things, the boundary
conditions. This is not meant to be a replacement for 
constraint-preserving boundary conditions but, instead, complementary
to them.

\subsection{Numerically stable but with fast growing errors simulations}

Evolutions of the PG spacetime with equations 
(\ref{adot},\ref{kadot},\ref{kbdot})  can give rise to errors that grow fast in time, even 
 with a stable numerical method. 

Figure 1 shows $N_c$ vs. time for evolutions of PG initial
data, at different resolutions. Typical numerical parameters are
chosen: Courant factor $\lambda = 0.25$, inner and outer boundaries at
$r_i=M, r_o=40M$, respectively, and dissipation parameter $\sigma =
0.1$. The constraints converge to zero
with resolution, but at fixed resolution they grow very fast in
time. This is not a peculiarity of this formulation and this numerical
method but, indeed, is typical for free evolutions of Einstein's
equations in the strong field regime.

One could attempt to modify the numerics in order to
minimize the growth of the constraints, but I do not pursue this approach
 here.
\begin{center}
\begin{figure}
\epsfig{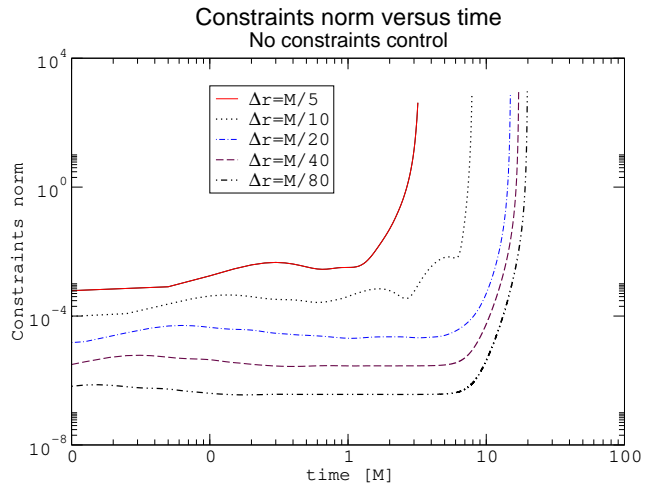}
\caption{Evolution of the Painleve\'e-Gullstrand initial data, without
 controlling the constraints ($\mu=0$). $N_c$ converges to
 zero when resolution is increased, 
but at fixed resolution it grows fast in time, making the code crash
 around $t=10M$.}
 \label{unstable}
\end{figure}
\end{center}

An interesting feature of the instabilities of the simulations shown
in Figure 1 is that the norm growth is initially negative, but it
becomes positive after a single time step, triggering the instability, 
see (\ref{fluc}) (there, as in the rest of the paper,
$\dot{N}_c$ is computed through the semidiscrete expression
(\ref{split})). 
\begin{center}
\begin{equation}
\begin{tabular}{||l|l|l|l||}\hline
$\Delta r$ & $\Delta t$ & $\dot{N}_c(0)$ & $\dot{N}_c(\Delta t)$ \\ \hline 
$M/5 $&  $5.0\times 10^{-2}$& $-5.01\times 10^{-3}$ & $ 1.47\times
  10^{-2}$ \\\hline
$M/10 $&  $2.5\times 10^{-2}$& $-1.81\times 10^{-3}$ & $
    3.85\times 10^{-3} $ \\\hline 
$M/20 $&  $1.25\times 10^{-2}$ & $-5.39\times 10^{-4}$  & $ 9.54\times 10^{-4}$ \\\hline 
$M/40 $&  $6.25\times 10^{-3}$ & $ -1.46\times 10^{-4}$  & $ 2.34\times 10^{-4}$ \\\hline 
$M/80 $&  $3.12\times 10^{-3}$ & $ -3.82\times 10^{-5}$  & $ 5.76\times 10^{-5}$ \\\hline 
\end{tabular}
\label{fluc}
\end{equation}
\end{center}
A static choice of $\mu$ cannot account for these fluctuations. 
In particular, choosing $\mu$ using only the 
growth rate around the background solution, which in this case would
be the initial data,  and not the errors introduced during
numerical integration, in general will not be able to 
 correct these kind of fluctuations in $\dot{N}_c$.

\subsection{Strictly enforcing semidiscrete norm preservation}
I now discuss some numerical experiments using a dynamically
calculated $\mu(t)$ such that $\dot{N_c}=0$ in the semidiscrete case. 
For a chosen resolution, $\mu$ is defined at each time step by 
\begin{equation}
\mu(n\Delta t) =\left\{
\begin{array}{l}
-I^{(1)}(n\Delta t)/I^{(2)}(n\Delta t)\;, \; \mbox{ if }I^{(2)}(n\Delta
t) \neq 0 \\
\mu\left((n-1)\Delta t\right) \;, \; \mbox{  if } I^{(2)}(n\Delta
t) = 0 \mbox { and } {n\geq 1} \\
 0 \; , \;\;\mbox{  if } I^{(2)}(n\Delta
t) = 0 \mbox { and } {n=0} 
\end{array}
\right. \label{strict_eqs}
\end{equation}
and $\mu(n\Delta t)$ is in turn used to compute the RHS needed to 
advance $u\left(n \Delta t\right)$. 

\begin{center}
\begin{figure}
\epsfig{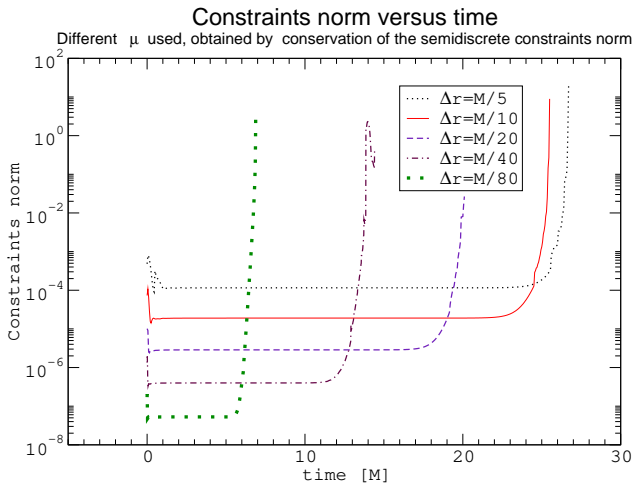} 
\caption{Strict conservation of the semidiscrete norm ($\dot{N}_c=0$), 
using different resolutions 
to define $\mu(t)$, i.e. $\mu$ is dynamically computed at each
resolution. The fact that the errors increase when resolution is
increased strongly suggest that defining $\mu$ this way leads to an ill-posed
problem. The appropriate way of doing a convergence test is shown in
Figure \ref{strict_convergence}.}
\label{strict_energy}
\end{figure}
\end{center}

\begin{center}
\begin{figure}
\epsfig{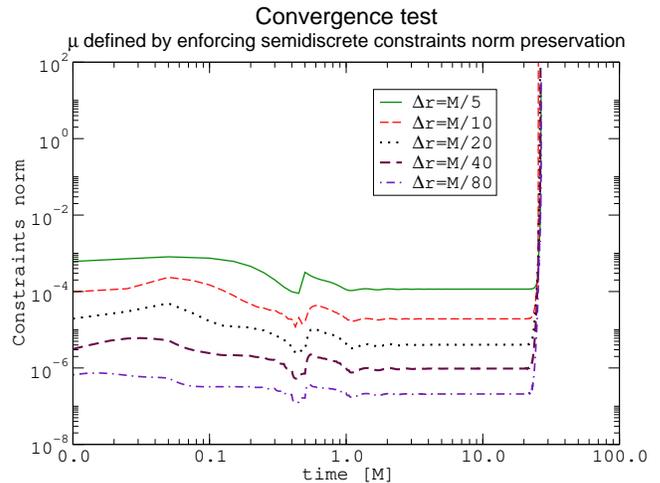} 
\caption{Strict conservation of the semidiscrete norm, defining
$\mu$ by a $\Delta r=M/5$ run and then fixing $\mu $ 
 for all other runs.}
\label{strict_convergence}
\end{figure}
\end{center}

Figure \ref{strict_energy} shows the resulting norm as a function
of time, for evolutions enforcing $\dot{N}_c=0$ at different resolutions. The code
blows up rather fast,
and this happens at earlier times when resolution is increased. 
This last feature should not be seen as an indication of ill-posedness of the method,
since Figure \ref{strict_energy} should {\em
  not} be seen as a convergence test. For such a test one should 
dynamically define $\mu(t)$ with a given resolution and
then keep that $\mu$ fixed for all other runs.  
Figure \ref{strict_convergence} shows the result of
doing so, defining $\mu(t)$ through a run with $\Delta r=M/5$ and then keeping
it fixed for a convergence test \footnote{Since $\mu$ is obtained for 
the coarsest resolution used in the convergence test,
some interpolation procedure is needed in order to have this function
defined at intermediate time steps. The simple procedure I have followed
here is to define the needed intermediate values of $\mu$ between time
steps $n$ and $n+1$ just as $\mu(n)$, though better
(e.g. trigonometric) interpolation could be used. Another option would
be to define $\mu$ through the finest resolution that is going to be
used for the convergence test and later coarsen the obtained discrete
$\mu$ function.}. As expected, and in contrast to Figure  
\ref{strict_energy}, 
the norm does decrease as the resolution is increased. Figure
\ref{strict_energy}
strongly suggests that if one insisted in 
dynamically defining $\mu$ at each resolution, instead of fixing it at
a single resolution, one would end up with an ill-posed problem.  On the other hand,
fixing $\mu$ leads to a strongly hyperbolic problem and no
convergence problems appear. 

\begin{center}
\begin{figure}
\epsfig{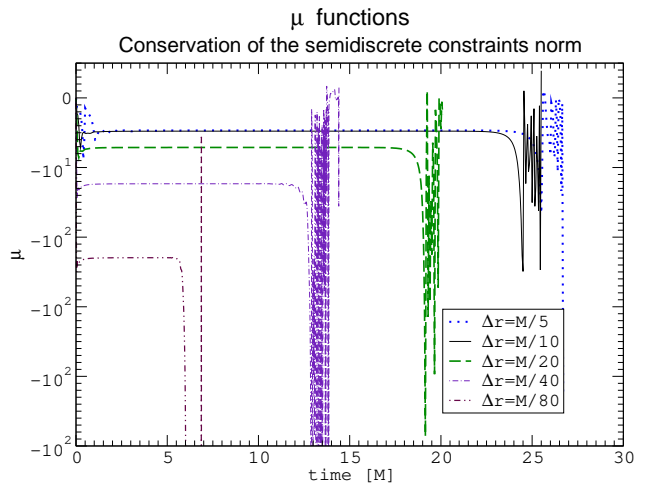}
\caption{This figure shows $\mu(t)$ for the runs of Figure 2.}
\label{strict_mu}
\end{figure}
\end{center}

Considering these runs with $\dot{N}_c=0$, it appears that enforcing 
semidiscrete norm preservation is ``too rigid'' (recall the discussion
at the beginning of
Section III). Figure \ref{strict_mu} shows the $\mu$ functions obtained through
the runs of Figure \ref{strict_energy}. Notice that when increasing resolution the
$\mu$ that is needed in order to preserve the norm increases quite
fast in absolute value. For example, with $\Delta r=M/80$ the
initial values of $\mu$ are of order $10^2$. Since $\mu$ appears in
the principal part of the equations, having a large value implies the
need of a small Courant factor to follow the simulation; while Figures
\ref{strict_energy}  and
\ref{strict_mu}, on the other hand, where obtained using the same Courant factor
when changing resolution. A Fourier decomposition of the
numerical solution, or an explicit von-Neumann analysis of a linear,
constant coefficient problem could give further insight into this question.

A second possibility would be to perform a fully discrete analysis (as
opposed to the semidiscrete one used in this paper), and to enforce
strict conservation of the fully discrete norm, but these issues are not pursued
here. 

\subsection{Dynamically achieving a tolerance value}
I now set a tolerance value $T$, and choose $\mu$ such that $N_c=T$
 after a given number of timesteps. In doing so I use 
the semidiscrete expressions (\ref{a},\ref{mu}). If  
 $I^{(2)}$ is zero, $\mu$ will be copied from its previous value, or set to zero if this
happens in the first iteration. 
\begin{center}
\begin{figure}
\epsfig{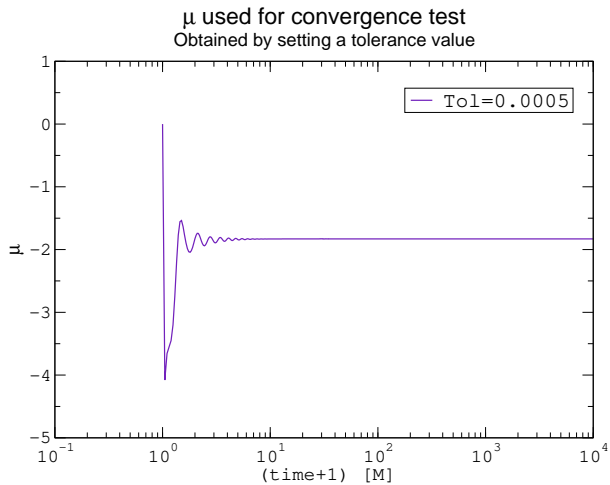}
\caption{This figure shows $\mu$ calculated by setting a tolerance
  value, 
  $T=5\times 10^{-4}$, and requiring the semidiscrete norm to return
  to this value after every timestep. $\mu$ is defined using $\Delta
  r=M/5$, for which the initial, truncation
  value of the norm is $N_c=4.8\times 10^{-4}$.}
\label{mu_conv}
\end{figure}
\end{center}
Figure \ref{mu_conv} shows results with a dynamically defined $\mu$,
using the same 
resolution, domain, Courant and dissipation 
factors as before. $n_a$ is chosen to be one, 
corresponding to the norm returning to the specified tolerance value after
every time step. 

The discretized constraints are, of course, 
not zero initially, even when the initial data is analytic, 
because computing them involves finite
differencing the initial data. The initial value for the norm for 
this resolution is $N_c=4.8\times
10^{-4}$. Therefore I choose a tolerance value $T=5\times 10^{-4}$, to
keep the constraints close to its initial value. 

\begin{center}
\begin{figure}
\epsfig{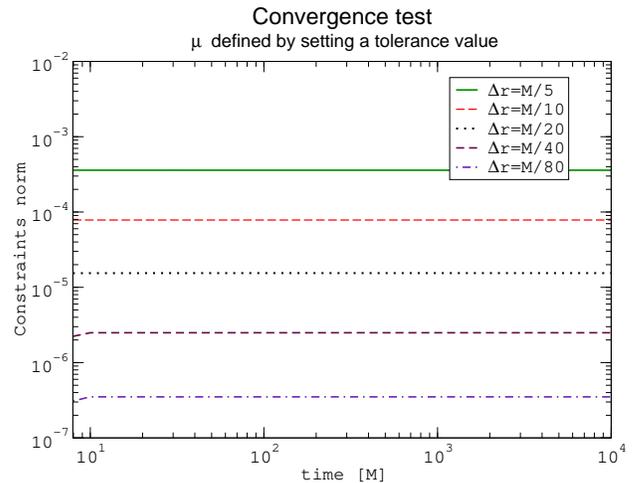}
\caption{A convergence test done with  $\mu$ shown in
  Figure \ref{mu_conv}. Compare the lifetime of these simulations with
  those of   Figure \ref{unstable}.}
\label{convergence}
\end{figure}
\end{center}

Figure \ref{convergence} shows the resulting norms for a convergence test performed with
the same resolutions used in Figure \ref{unstable}, and the $\mu$ shown in Figure
\ref{mu_conv}. The code runs for $10^4M$ without any sign of
instabilities, even with the coarse
resolution that was used to define $\mu$. 

\begin{center}
\begin{figure}
\epsfig{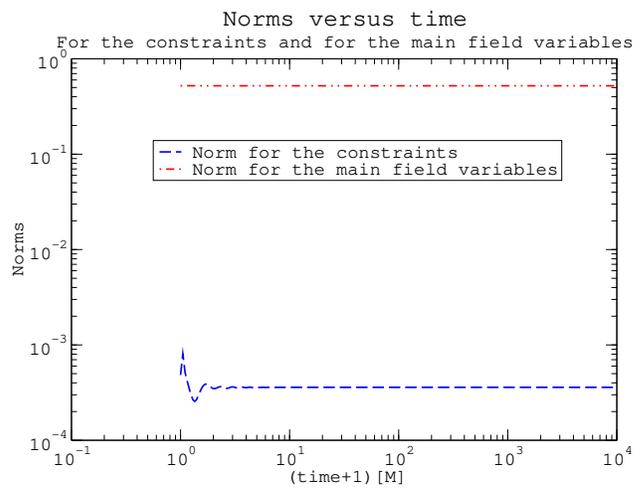}
\caption{Norms associated with the constraints and with the main
  field variables, for the run used to obtain $\mu$ shown in Figure \ref{mu_conv}.}
\label{norms2}
\end{figure}
\end{center}

It could happen that one controls the constraints at the price of
introducing other errors. This does not happen, at least in the
numerical experiments here considered. In these experiments, not only
does the norm associated with the constraints remains close
to its initial, truncation value, but the same happens with the norm
associated with the main field variables, $N=\int
(a^2+K_a^2+K_b^2)$. Since the exact solution is stationary, $N$ being
close to its initial value means that the method does not introduce
new errors. Figure \ref{norms2} shows $N_c(t)$ and $N(t)$ for
the run used to define $\mu$. 

\subsubsection{Dependence of $\mu$ on the tolerance value}
In the previous examples I chose the tolerance value, $T$, to be roughly the initial,
truncation error for $N_c$. Although this seems the natural thing to do, I
will now discuss sensitivity of the stability method with respect to the
chosen value of $T$.
\begin{center}
\begin{figure}
\epsfig{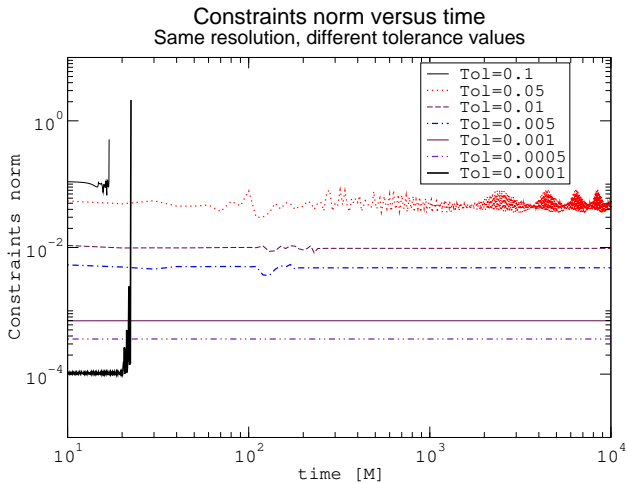}
\caption{This figure shows the sensitivity of the method to the chosen
  tolerance value $T$. Values too large (a couple of order of
  magnitudes bigger than the initial truncation error $N_c(0)$), or
  smaller than
  $N_c(0)$, lead to instabilities. Otherwise, no fine tuning of
  $T$ is required.}
\label{fenergy}
\end{figure}
\end{center}
Figure \ref{fenergy} shows results when choosing different
tolerance values to define $\mu$, but keeping all other numerical parameters
(in particular, the resolution, i.e. Figure \ref{fenergy} {\em is not} a
convergence test) fixed. When $T$ is much larger than the 
initial truncation value of $N_c$, large errors allow  triggering of nonlinear
instabilities. On the other hand, values of $T$ considerably smaller than the initial truncation one 
forces $\mu$ to change very fast,
and to large values. In either case the code crashes.  
The $\mu$ functions for two such cases are shown in
Figure \ref{mu_bad}. 
\begin{center}
\begin{figure}
\epsfig{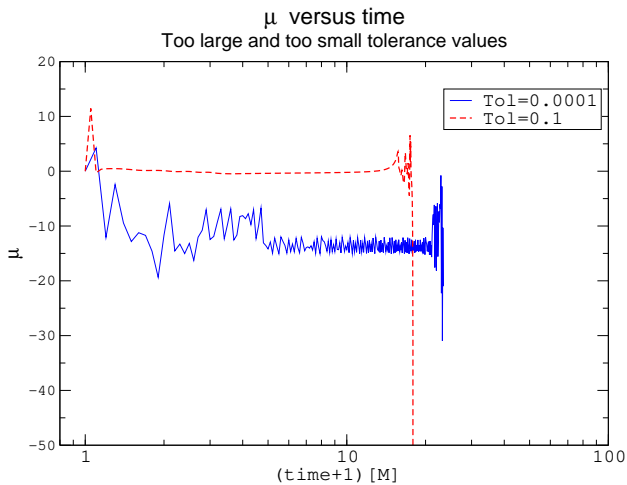} 
\caption{$\mu$ for two unstable cases, with too large 
  ($T=10^{-1}$) and too small tolerance values ($T=10^{-4}$).}
\label{mu_bad}
\end{figure}
\end{center}

\begin{center}
\begin{figure}
\epsfig{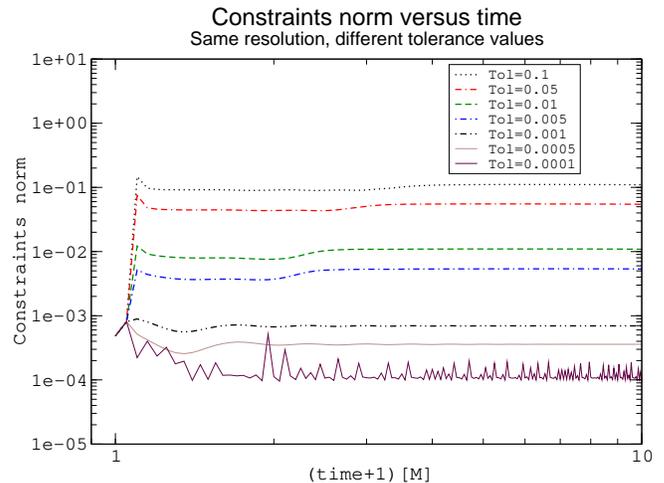} 
\caption{The figure shows details of plot \ref{fenergy}, near $t=0$. 
The norms in this experiment are the same in the
  initial data {\em and} first time step, as explained in
  the body of the paper.}
\label{energy_zoom}
\end{figure}
\end{center}

Figure \ref{energy_zoom} shows the details of Figure \ref{fenergy} near
$t=0$. Notice that not only all the runs begin with the same (truncation) value for
$N_c$ (this is expected, since the same resolution is used), but
$N_c$ is also the same
in the very first timestep. The reason for this is that $\mu =0$ initially for all
the runs, because $I^{(2)}=0$ in the initial data. This is a
coincidence of the initial data being evolved in this example: the 
only discretization error in the
computation of the Hamiltonian constraint in this model comes from the
finite differencing of $a$, which in this example is initially $1$ for all grid
points. 

\subsection{Intentional constraint violations}
\begin{center}
\begin{figure}
\epsfig{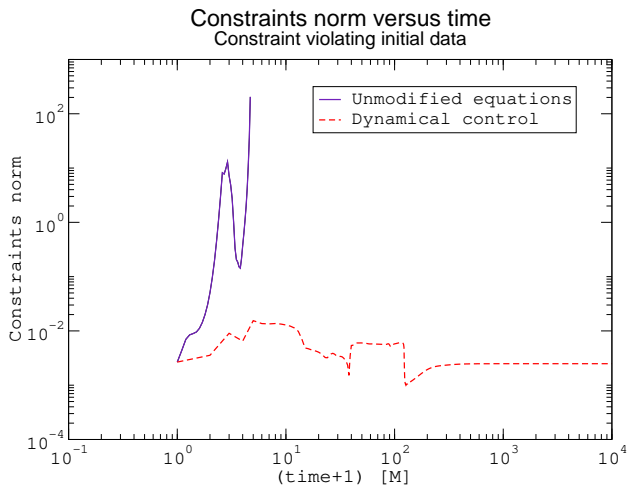}
\caption{The norm $N_c$, with and without control of the constraints, for
  highly perturbed initial data, at a resolution of $\Delta r=M/5$. The initial, truncation error for
  $N_c$ for this resolution is $N_c(0)=2.66\times 10^{-3}$, and the
  tolerance value is set to $T=5\times 10^{-3}$.}
\label{violation}
\end{figure}
\end{center}

\begin{center}
\begin{figure}
\epsfig{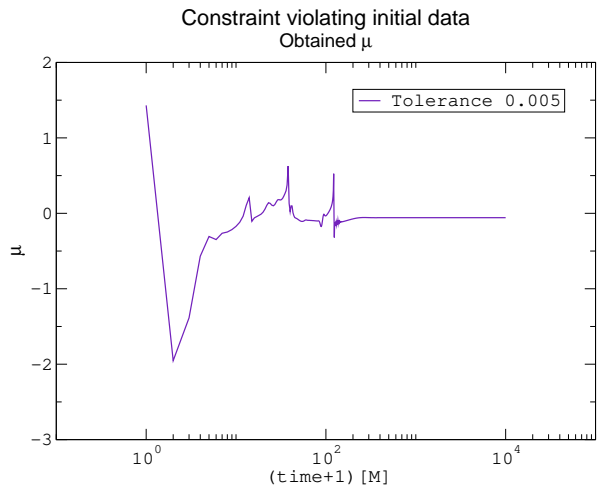}
\caption{$\mu$ for the plot of Figure \ref{violation}
  with dynamical controlled constraints. After some time $\mu$
  settles down to the value $\mu = -5.8 \times 10^{-3}$.}
\label{violation_mu}
\end{figure}
\end{center}

As a final numerical example, now I consider initial data 
that violates the constraints at
the continuum. The equations still use the exact lapse and  
area-locking shift given  by Eq.(\ref{gauge}), but now the initial 
 data are given by
\begin{eqnarray}
a&=&\left(1+s_1e^{-(r-r_0)^2}\right) \\
K_a &=& -\frac{\beta }{2r}
\left(1+s_2e^{-(r-r_0)^2}\right)  \\
K_b &=& \frac{\beta}{r}=\left(1+s_3e^{-(r-r_0)^2}\right)
\end{eqnarray}
When $s_i=0$, this data reduces to that of the PG
black hole. The constants $s_i$ are used to introduce an ``asymmetry'' in the 
perturbations of the PG initial data. The perturbations are rather
large, with $s_1$ or order unity. In this example, 
$s_1=1,s_2=0.5,s_3=0.7$. When $\mu = 0$ the code crashes before $t=4M$ 
at a resolution $\Delta
r=M/5$.  At this resolution $N_c$ for the initial data is  
$N_c(0)=2.66\times 10^{-3}$ (one order of magnitude larger 
than the associated truncation error for the case $s_i=0$, which for
this resolution is $N_c=4.8\times 10^{-4}$ ). Thus, I choose a
tolerance value of $T=5\times 10^{-3}$, $n_1=10$ in order to avoid
fast variations,  and 
rerun, dynamically obtaining $\mu(t)$. Results for the unmodified 
equations ($\mu =0$), and with a dynamical  $\mu$ are shown in Figures  
\ref{violation}, \ref{violation_mu}. The difference in stability is quite striking,
especially considering that the perturbation of the PG 
initial data is so large. 

\section{Remarks}

There is a lot of interest in finding formulations
of Einstein's equations with good stability properties off of the
constraint surface. 
So far, most, if not all, of the work
has been concentrated in stability properties around known
solutions. The purpose of this paper has been to introduce new ideas
that hopefully will be helpful in generic evolutions, where one
has no a priori knowledge of the solution. This is one fundamental
difference with previous approaches. 

Another difference is that
previous work has concentrated on stability properties at
the continuum. Although this is an important issue, controlling
constraint violating modes introduced by the numerical method is  
 also crucial. Thus, this proposal 
 aims to control discrete constraint violations introduced
both by the underlying formulation of equations at the continuum, and
whatever numerical method one has chosen. 

Sometimes it seems that one would ideally want the constraints to decay exponentially to
zero \cite{lambda},\cite{fiske}. Although this
might be attractive at the continuum, in what concerns the discrete
constraints it might be better to maintain them  
around the initial, truncation error.  For example, beginning with 
 initial data corresponding to a stationary spacetime, it seems
difficult to imagine that by making the
constraints decrease in time through numerical integration (keeping
resolution fixed), one would end up
with a numerical solution that has smaller errors than the initial data, which was
computed by just numerically evaluating the exact solution at given
gridpoints. Thus, I have concentrated here on keeping 
the discrete constraints at some tolerance value, instead of forcing them
to decay to zero. However, there may be other scenarios where it is
preferable to make them decay to values smaller than the initial one,
and the method here presented allows for this as well. Along these
lines, an extension of the work presented here
is to use the evolution equations applied to initial
data off the constraint surface to produce solutions
of the constraints as an alternative to relaxing the constraints
themselves \footnote{L. Lehner, private communication}.

In some other approaches \cite{ls}, a norm for the main variables, say
$N(t)=\int _{S_t} \sum_i u_i^2$, is minimized with respect to
constraint violating perturbations (around a given, known
background). I could have here chosen to dynamically minimize this
quantity. One potential problem
with this is that it is not clear how much minimization makes the
solution closer to the ``exact'' one. For example, in the spherically
symmetric example here considered the time derivative of the norm
for the main variables has the form
$$
\dot{N} = \tilde{I^{(1)}} + \mu \tilde{I^{(2)}}
$$
Since $\mu$ is completely free in this model,
one is able to make $\dot{N}$, for some chosen but arbitrary
resolution, as negative as one wants. What happens
then is that all the field variables decay to
zero (numerical experiments that I have done do confirm this). 
Clearly this is not what one wants, since the ``exact'' solution
is not identically zero. On the other hand, one could also attempt to keep
$N$ close to its initial, truncation value. However, this would be a
good idea only if the solution is stationary, since otherwise there
is no reason for the exact $N(t)$ to be close to $N(0)$ for all
times. Thus I have here considered a norm associated
with the constraints, since they are analytically zero, regardless of
the solution.

\section{Acknowledgments}

I would like to thank Oscar Bruno, Gioel Calabrese, Luis Lehner, David
Neilsen, Jorge Pullin, Oscar Reula, Olivier Sarbach, Ed Seidel and
Jonathan Thornburg for very helpful discussions, comments and
suggestions. This work was supported in part
by NSF grant PHY9800973, the Horace Hearne Jr. Institute for
Theoretical Physics, and Fundaci\'on Antorchas.

\end{document}